\documentclass[aps,prx,preprint,superscriptaddress]{revtex4}  
\usepackage[dvips]{graphicx}
\usepackage{latexsym}
\usepackage{natbib}

\begin{document}
\title{Cascade of failures in coupled network systems with multiple
  support-dependent relations}
 
\author{
Jia Shao$^{1}$, Sergey V. Buldyrev$^{2,1}$,
Shlomo Havlin$^{3}$ and H. Eugene Stanley$^{1}$
}
               
\affiliation{
$^1$Center for Polymer Studies and Department of Physics,
  Boston University, Boston, Massachusetts 02215, USA\\
 $^2$Department of Physics, Yeshiva University, 
  500 West 185th Street, New York, New York 10033, USA.\\
  $^3$Minerva Center and Department of Physics, Bar-Ilan University, 52900
Ramat-Gan, Israel\\
}

\date{ Printed: \today }

\begin{abstract}

  We study, both analytically and numerically, the cascade of failures
  in two coupled network systems A and B, where multiple
  support-dependent relations are randomly built between nodes of
  networks A and B. In our model we assume that each node in one
    network can function only if it has at least a single support node
    in the other network.  If both networks A and B are
  Erd\H{o}s-R\'enyi networks, A and B, with (i) sizes $N^A$ and $N^B$,
  (ii) average degrees $a$ and $b$, and (iii) $c^{AB}_0N^B$ support
  links from network A to B and $c^{BA}_0N^B$ support links from
  network B to A, we find that under random attack with removal of
  fractions $(1-R^A)N^A$ and $(1-R^B)N^B$ nodes respectively, the
  percolating giant components of both networks at the end of the
  cascading failures,
  $\mu^A_\infty$ and $\mu^B_\infty$, are given by the
  percolation laws
  $\mu^A_\infty=R^A[1-\exp{({-c^{BA}_0\mu^B_\infty})}][1-\exp{({-a\mu^A_\infty})}]$
  and
  $\mu^B_\infty=R^B[1-\exp{({-c^{AB}_0\mu^A_\infty})}][1-\exp{({-b\mu^B_\infty})}]$.
  In the limit of $c^{BA}_0 \to \infty$ and $c^{AB}_0 \to \infty$,
  both networks become independent, and the giant components are
  equivalent to a random attack on a single Erd\H{o}s-R\'enyi
  network. We also test our theory on two coupled scale-free networks,
  and find good agreement with the simulations.
\end{abstract}

\maketitle
\section{Introduction}

In recent years, there has been extensive effort to study and
understand the properties of complex networks.
Research has mainly focused on properties of single networks which do
not interact or depend on other networks \cite{watts, bara, vespig,
  mendes, cohen, newman, bocalletti, havlinbook,
  caldarelli, barthelemy,newmanbook, callaway, shaoprl,
  lidia}. Recently, the robustness of two interdependent coupled networks
has been studied \cite{buldyrev,parshani}. In interdependent networks,
the failures of nodes in one network, A, will cause failures of dependent
nodes in the other network, B, and {\it vice versa}.  This process occurs
recursively, and leads to a cascade of failures. It has been shown
both analytically and numerically that the robustness of two
interdependent networks is significantly lower compared to that of a
single network \cite{buldyrev}. Furthermore, the percolation
transition in coupled networks is first order compared
to the known second order transition in a single network
\cite{buldyrev,parshani}.

Previous studies on two interdependent coupled networks are restricted
by the condition, that to function each node in network
A depends on one and only one node in network B
and {\it vice versa} \cite{buldyrev}. However, in the real world, this
assumption may not be valid. A single node in network A may depend on
more than one node in network B and will function as long as one of the
support nodes in network B is still connected. Similarly, a node
in network B may depend on more than one support nodes in network A. As long
as one of the support nodes functions, the node in network B will also
function. 

Examples of such systems include the coupled power grid network and
the communication network which controls the power grid, where both
networks depend on each other. In general, one power station provides
power to more than one communication stations, and one communication
station controls more than one power stations.  As long as a
communication station can obtain power from one power station, it can
function properly. On the other hand, one communication station is
sufficient in sending control signals which make one power station
function properly. However without any power, the communication
station will fail, and without control the power station will stop
working. Indeed in the 2003, due to failure of some power stations in
Italy, the communication control system was damaged.  This damage
caused further fragmentation of the power grid, which finally led to a
blackout in a sinificant part of Italy.



 Under random attack, which is characterized by random removal of nodes in
 one or both networks, the coupled network systems demonstrate
 significantly different behavior from that of a single network
 \cite{buldyrev}. The failures of nodes in network A can lead to the
 failures of dependent nodes in network B, and the failures of nodes in
 network B can produce a feedback on network A leading to further
 failures in network A. This process can occur recursively and can lead
 to a cascade of failures.

 We provide a theoretical framework for understanding the robustness
 of interdependent networks with a random number of support and
 dependent relationships. Our theory agrees well with the numerical
 simulations of several model network systems, including coupled
 Erd\H{o}s-R\'enyi (ER) \cite{er} and coupled scale-free (SF)
 \cite{bara} networks. Our work extends previous works on
   coupled networks \cite{buldyrev,parshani} from {\it one-to-one}
   dependent-support relation to {\it multiple} Poissonian dependent-support
   relation.  Our model can help to further understand real-life
   coupled network systems, where complex dependence-support relations
 may exist.


We define the stable state to be the state when the cascade of
failures ends. We show that for two coupled ER networks the giant
components of both networks in the stable state follow a simple law,
which is equivalent to random percolation of a single network in the
limit of a large number of support links.  Our theory is relevant to a
broad class of real-world interdependent network systems.

The paper is organized as follows. In Sec. II, we explain the model of
the cascade of failures with random support-dependent relations. In
Sec. III, we derive analytically the process of the failure
cascade. In Sec. IV, we present numerical tests on coupled ER and SF
networks.

\section{The Model}

We assume two networks A and B of sizes $N^A$ and $N^B$ and with given
degree distributions, $P^A(k)$ and $P^B(k)$, of ``intra-links''
connecting nodes in the same network (Fig. 1). The dependency relation
is represented by a link connecting the support node in one network
and the dependent node in the other network (``inter-links'').  The
inter-links between network A and network B are random and
uni-directional. Initially (stage $n=0$ of the failure cascade), there
are $c^{BA}_0N^A$ inter-links distributed randomly from nodes in
network B to nodes in network A, representing the dependencies of node
in network A on nodes in network B. Similarly, there are $c^{AB}_0N^B$
inter-links from nodes in network A to nodes in network B,
representing dependencies of nodes in network B on nodes in network A.
The dependent-support relations are random i.e., for each inter-link,
the support node and dependent node in the two networks are chosen
randomly. $c^{BA}_0$ and $c^{AB}_0$ are the initial mean degrees of
the corresponding inter-links for networks A and B respectively.

In our model, one node in either network A or network B can have zero,
one or several support nodes in the other network. We assume that to
function each node in network A requires at least one support node in
network B, and {\it vice versa}. Nodes in each network which are not
connected to the giant components of that network by the intra-links,
and nodes without support inter-links, are considered to be not
functioning and regarded as failed nodes.

The attack on the network is represented by a random removal of
fraction $1-R^A$ nodes in network A and $1-R^B$ in network B, 
  where in general, $R^A \neq R^B$. The process of the cascade of 
failures is demonstrated in Fig. 1. With $N^A=N^B=7$, we show
the case of random removal of one node in network A ($R^A=6/7$) and
one node in network B ($R^B=6/7$). At each stage of the cascade of
  failures, both networks will experience further failures.  Without loss of 
generality, we assume the random attack on network A occurs before that on
network B. Thus, when we analyze the first stage of the cascade of
failures in network A, all the support inter-links from network B are
considered functioning. At each stage, the nodes which do not have any
support inter-links from the other network, and the nodes which are
separated from the giant component of the network, are considered to have
failed. This process will continue until no further node failure 
in either network occurs.


\begin{figure}
  \includegraphics[width=17cm,angle=0]{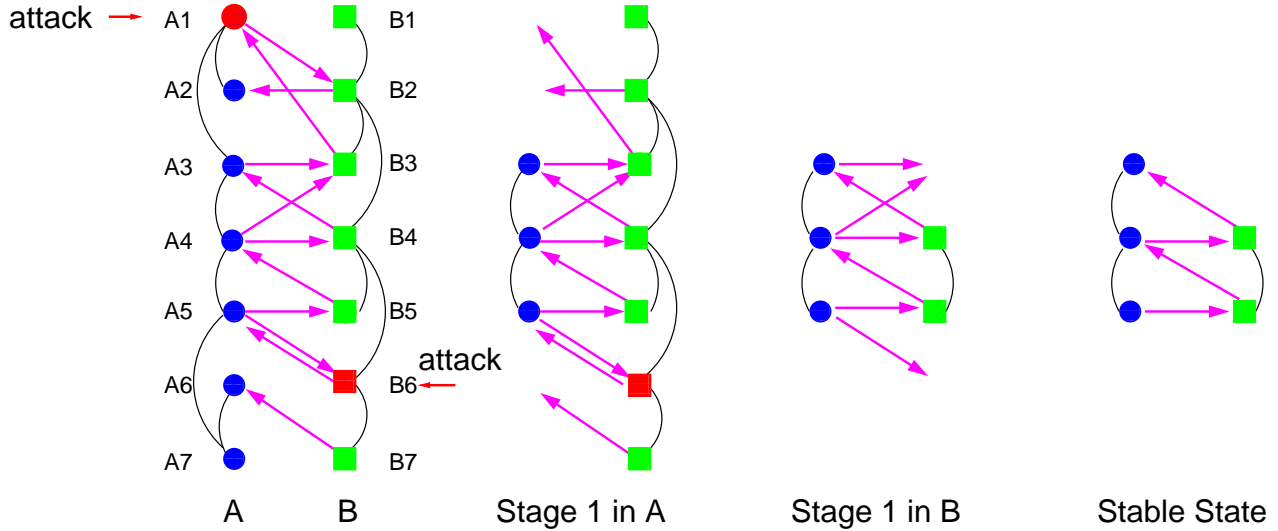}
  \caption{ (Color online) Demonstration of the stages of the cascade
    of failures in coupled network A and B of size $N^A=N^B=7$.
    Curves represent intra-links within the network, while arrows
    (directed links) represent the inter-links connecting a support
    node in one network to the dependent node in the other
    network. Among the total 12 directed links, half of them
    $c^{BA}_0N^A=6$ represent the support from nodes in network B to
    nodes in network A (arrow from nodes in network B to nodes in
    network A).  The rest $c^{AB}_0N^B=6$ links represent the support
    from nodes in network A to nodes in network B. The
    support-dependent relations between nodes in network A and network
    B are random. Initially, the attack is on node $A_1$ (shown in
    red) in network A and node $B_6$ (shown in red) in network B.  The
    failed nodes are removed from the plot.  In the first stage of the
    cascade of failures in network A, $A_1$ fails because of removal,
    node $A_7$ fails because of it has no support inter-links, $A_2$
    and $A_6$ fail because of separation from the giant component of
    network A.  All the failed nodes will lead to failures of
    inter-links starting from them.  Since we assume that the attack
    on network A occurs before that on network B, the inter-link from
    $B_6$ to $A_5$ is considered to be functional.  In the first stage
    of the cascade of failures in network B, we first remove
    inter-links connecting network B to non-giant-component nodes in
    network A ($B_3$ to $A_1$, $B_2$ to $A_2$, $B_7$ to $A_7$). Next,
    node $B_6$ fails because of the attack and nodes $B_1$, $B_2$ and
    $B_7$ also fail because of no support. $B_3$ fails because it
    becomes separated from the giant component (nodes $B_4$ and $B_5$)
    of network B.  Finally, after removing inter-links connecting from
    network A to non-giant-component nodes in network B ($A_3$ to
    $B_3$, $A_4$ to $B_3$ and $A_5$ to $B_6$), the coupled network
    system reaches a stable state after one step in the cascade of
    failures, since all nodes in both giant components are connected
    and each node have at least one support node from the other
    network.  }
\label{fig1}
\end{figure}
 
\section{Analytical Solution}


The stable state of the two stable connected giant components in both
networks are usually reached after several stages in the cascade of
failures.
We define $c^{AB}_n$ and $c^{BA}_n$ as the average number of support
inter-links remaining in the stage $n$ of the cascade of failures of
network A and network B respectively.

In each stage of the cascade of failures, we analyze first network A
then network B. Such a procedure does not have any affect on the final
result of the failure cascade.  While considering network A at stage
$n$, we assume all the inter-links ($c^{BA}_{n-1}$) from network B to
network A of the previous stage are working. When considering network
B at stage $n$, we use the updated $c^{AB}_{n}$ after the $n$th stage
of network A.


Since, in our model we randomly set up $c^{AB}_0N^B$ inter-links from
network A to network B and $c^{BA}_0N^A$ inter-links from network B to
network A, the degree distributions of inter-links of both networks
follow Poisson statistics. There exist nodes in network A and network
B without support initially (nodes $A_7$ and $B_1$ in Fig.1). These
nodes are regarded as failed and need to be considered in addition to
the random removal of nodes.
 
At stage 1, after removal of a fraction $1-R^A$ of nodes in
network A, taking into account also the nodes with zero 
inter-links at stage 0, the result is equivalent 
to a random removal of a fraction $1-p^A_1(R^A,c^{BA}_0)$ of
nodes in network A.  The giant component of network A will constitute
a fraction $g^A_1(p^A_1,c^{BA}_0)$ of the remaining $p^A_1N^A$ nodes of
network A. Thus, the fraction $\mu^A_1$ of the giant component with
respect to the $N^A$, the original size of network A, is
\begin{equation}\label{N1}
\mu^A_1=p^A_1(R^A,c^{BA}_0)g^A_1(p^A_1,c^{BA}_0).
\end{equation}
Since we assume that all support links from network B exist, we use $c^{BA}_0$.

The failures of $(1-\mu^A_1)N^A$ nodes in network A will lead to
failures of support inter-links from network A to network B. The working
support inter-links from network A to network B, $c^{AB}$, will
decrease from $c^{AB}_0$ to $c^{AB}_1(\mu^A_1)$.

Next, at the first stage of cascade of failures in network B, a
fraction $1-R^B$ of nodes in network B will fail because of the
initial attack. Combining this fraction with the fraction of nodes in network B
having zero degree of support links after the first stage of
  cascade of failures in network A, the joint effect is equivalent to
a random removal of a fraction $1-p^B_1(R^B,c^{AB}_1)$ of nodes in
network B.
The giant component of network B will constitute a fraction
$g^B_1(p^B_1,c^{AB}_1)$ of $p^B_1N^B$ nodes.  Thus the fraction of the giant
component of network B after the 1st stage of cascade of failures is
\begin{equation}\label{N2}
\mu^B_1=p^B_1(R^B,c^{AB}_1)g^B_1(p^B_1,c^{AB}_1).
\end{equation}
The number of support links $c^{BA}$ then will be reduced from
$c^{BA}_0$ to $c^{BA}_1(\mu^B_1)$.

The 2nd stage of the cascade of failures in network A is equivalent to
the first stage in network A with the updated $c^{BA}_1$ replacing
$c^{BA}_0$. Accordingly, the 2nd stage of cascade failures in network
B is equivalent to the first stage in network B with the updated
$c^{AB}_2$. During the cascade of failures, $c^{AB}_n$
and $c^{BA}_n$ decrease as $n$ increases.


In general, the $n$th stage of the cascade in network A is equivalent to
the 1st stage of the cascade in network A with $c^{BA}_0$ replaced by
$c^{BA}_{n-1}$. Similarly, the $n$th stage of the cascade in network B is
equivalent to the 1st stage of the cascade in network B with $c^{AB}_1$
replaced $c^{AB}_{n}$. The general forms of the giant components
  of both networks at stage $n$ of the cascade of failures can be
  expressed as
\begin{equation}\label{general}
 \left\{%
\begin{array}{ll}
\mu^A_n=p^A_n(R^A,c^{BA}_{n-1})g^A_n(p^A_n,c^{BA}_{n-1}), \\
\mu^B_n=p^B_n(R^B,c^{AB}_n)g^B_n(p^B_n,c^{AB}_n),
\end{array}%
 \right.
\end{equation}
where $p^A_n$ and $p^B_n$ are the equivalent fractions of nodes in
network A and network B respectively after random removal, and $g^A_n$
and $g^B_n$ are the fractions of the giant components in the remaining
$p^A_n$ and $p^B_n$ fraction of nodes.  The key to the analytical solution
of this process is to find the way how $c^{AB}_n$ and $c^{BA}_n$
decrease with the cascade stage $n$.




Next, we will use the apparatus of generating functions \cite{mnewman}
to derive the analytical forms of $\mu^A_n$ and $\mu^B_n$, $c^{AB}_n$ and
$c^{BA}_n$.
The generating functions of the degree distribution $P^A(k)$ of
network A and $P^B(k)$ of network B are
\begin{equation}\label{GA0}
 \left\{%
\begin{array}{ll}
G_{A0}(x)\equiv \Sigma_{k=0}^{\infty}P^{A}(k)x^k, \\
G_{B0}(x)\equiv \Sigma_{k=0}^{\infty}P^{B}(k)x^k.
\end{array}%
 \right.
\end{equation}
Analogously, the generating functions of the underlying branching
processes are
\begin{equation}\label{GA1}
  \left\{%
\begin{array}{ll}
G_{A1}(x)\equiv G_{A0}'(x)/G_{A0}'(1), \\
G_{B1}(x)\equiv G_{B0}'(x)/G_{B0}'(1).
\end{array}%
 \right.
\end{equation}
After random removal of a fraction $1-p$ of nodes, the remaining $p$
fraction of the network will have different degree distribution. The
new generation functions $G_0$ and $G_1$ will be \cite{shao1,shao2}
\begin{equation}\label{GA0n}
   \left\{%
  \begin{array}{ll}
G_{A0}(x,p)=G_{A0}(1-p(1-x)), \\
G_{B0}(x,p)=G_{B0}(1-p(1-x)), \\
G_{A1}(x,p)\equiv G_{A1}(1-p(1-x)), \\
G_{B1}(x,p)\equiv G_{B1}(1-p(1-x)).
 \end{array}%
 \right.
\end{equation}

Randomly connecting $c^{BA}N^A$ support links from network B to
network A, the degree distribution of the inter-links in network A 
follows a Poisson distribution with average degree $c^{BA}$
\begin{equation}\label{PC}
\tilde{P}^A(k)=\frac{[c^{BA}]^k}{k !}e^{-c^{BA}}.
\end{equation}
Similar, for network B,
\begin{equation}\label{PC}
\tilde{P}^B(k)=\frac{[c^{AB}]^k}{k !}e^{-c^{AB}}.
\end{equation}
During the process of the cascade of failures, because the
support-dependent relations are uncorrelated with the network properties
of network A and network B, $\tilde{P}^A(k)$ and
$\tilde{P}^B(k)$ will remain Poisson distributions with the new
$c^{BA}_n$ and $c^{AB}_n$, which decrease as $n$ increases.

Initially, there will be a fraction $\tilde{P}^A(0)=e^{-c^{BA}}$
of nodes in network A which do not have any support links
from network B. Since the attack on the $1-R^A$ fraction of nodes from
network A is random, there will be overlap between the attack and
the initially not working nodes (without support links) in network
A. The joint effect is equivalent to a random removal of a fraction
$1-R^A(1-e^{-c^{BA}})$ of nodes in network A.  What happens in
network B is similar to network A with $R^A$ replaced by $R^B$ and
$c^{BA}$ replaced by $c^{AB}$. Thus
\begin{equation}\label{p1}
 \left\{%
\begin{array}{ll}
p^{A}_n= R^A(1-e^{-c^{BA}_{n-1}}),\\
p^{B}_n= R^B(1-e^{-c^{AB}_n}),
 \end{array}%
 \right.
\end{equation}
where $c^{BA}_{n-1}$ and $c^{AB}_n$ are average degree of inter-links 
of network A and network B at the end of stage $n-1$ and $n$ respectively.


According to the results on single networks \cite{shao1, shao2}, after
random removal of a fraction $1-p^A$ (or $1-p^B$) of nodes, the
fractions of nodes that belong to the giant components of the
remaining network A or network B, which have $p^A_n$ and $p^B_n$
fractions of nodes respectively, are
\begin{eqnarray}\label{giAB}
 \left\{%
\begin{array}{ll}
g^A(p^A_n)=1-G_{A0}(f^A_n,p^A_n),\\
g^B(p^B_n)=1-G_{B0}(f^B_n,p^B_n).
 \end{array}%
 \right.
\end{eqnarray}
where $f^A_n$ and $f^B_n$ satisfy transcendental equations
\begin{eqnarray}\label{fAB}
  \left\{%
\begin{array}{ll}
f^A_n=G_{A1}(f^A_n,p^A_n),\\
f^B_n=G_{B1}(f^B_n,p^B_n).
  \end{array}%
 \right.
\end{eqnarray}
Thus $\mu^A_n$ and $\mu^B_n$, the fractions relative to their original
sizes of giant components of network A and network B
\cite{shao1,shao2} are
\begin{eqnarray}\label{muAB}
  \left\{%
\begin{array}{ll}
\mu^A_n=p^A_ng^A(p^A_n),\\
\mu^B_n=p^B_ng^B(p^B_n).
   \end{array}%
 \right.
\end{eqnarray}
Accordingly, $c^{AB}$ and $c^{BA}$ follow the relations
\begin{eqnarray}\label{cAB}
  \left\{%
\begin{array}{ll}
c^{AB}_n=c^{AB}_0\mu^A_n,\\           
c^{BA}_n=c^{BA}_0\mu^B_n.          
   \end{array}%
 \right.
\end{eqnarray}




When the cascade of failures process stops, $f^A_n$, $f^B_n$, $p^A_n$,
$p^B_n$, $c^{AB}_n$, $c^{BA}_n$, $\mu^A_n$ and $\mu^B_n$ all reach the
constant values, $f^A_\infty$, $f^B_\infty$, $p^A_\infty$,
$p^B_\infty$, $c^{AB}_\infty$, $c^{BA}_\infty$, $\mu^A_\infty$ and
$\mu^B_\infty$ respectively. In principle, these final values can be
found from the set of equations:
\begin{eqnarray}\label{group}
\left\{%
\begin{array}{ll}
f^A_\infty=G_{A1}(f^A_\infty,p^A_\infty),\\
f^B_\infty=G_{B1}(f^B_\infty,p^B_\infty),\\
p^A_\infty= R^A(1-e^{-c^{BA}_\infty}),\\
p^B_\infty= R^B(1-e^{-c^{AB}_\infty}), \\
c^{AB}_\infty=c^{AB}_0g^A(p^A_\infty)=c^{AB}_0[1-G_{A0}(f^A_\infty,p^A_\infty)],\\
c^{BA}_\infty=c^{BA}_0g^B(p^B_\infty)=c^{BA}_0[1-G_{B0}(f^B_\infty,p^B_\infty)],\\
\mu^A_\infty=p^A_\infty g^A(p^A_\infty),\\
\mu^B_\infty=p^B_\infty g^B(p^B_\infty).
  \end{array}%
 \right.
\end{eqnarray}

The functional forms of $G_{A1}$, $G_{B1}$, $G_{A0}$ and $G_{B0}$ can
be complicated, thus Eqs.(\ref{group}) can only be solved numerically
for most cases,  including coupled SF networks. However, for ER
networks, $G_0(x)$ and $G_1(x)$ have the same simple form
\cite{mnewman},
\begin{equation}\label{Eq.GER}
G_{0}(x)=G_{1}(x)=e^{\langle k \rangle(x-1)},
\end{equation}
where for network A, $\langle k\rangle=a$ and for network B, $\langle
k \rangle=b$.  Thus, the above process of the cascade of
failures can be significantly simplified.  Eqs.(\ref{giAB}) can be
reduced to
\begin{equation}\label{giABer}
  \left\{%
  \begin{array}{ll}
    g^A(p^A)=1-f^A,\\
    g^B(p^B)=1-f^B.
  \end{array}%
 \right.
\end{equation}

Excluding $p^A_\infty$, $p^B_\infty$, $c^{AB}_\infty$,
$c^{BA}_\infty$, $\mu^A_\infty$ and $\mu^B_\infty$ from
Eqs.(\ref{group}), for the stable state of two coupled ER networks, we
get a system of two equations with two remaining unknowns,
$f^A_\infty$ and $f^B_\infty$
\begin{equation}\label{fab0er}
   \left\{%
\begin{array}{ll}
f^A_\infty=G_{A1}(f^A_\infty,R^A(1-e^{-c^{BA}_\infty}))=e^{aR^A(f^A_\infty-1)(1-e^{-c^{BA}_\infty})}=e^{aR^A(f^A_\infty-1)(1-e^{c^{BA}_0(f^B_\infty-1)})},\\ 
f^B_\infty=G_{B1}(f^B_\infty,R^B(1-e^{-c^{AB}_\infty}))=e^{bR^B(f^B_\infty-1)(1-e^{-c^{AB}_\infty})}=e^{bR^B(f^B_\infty-1)(1-e^{c^{AB}_0(f^A_\infty-1)})}.
   \end{array}%
 \right. 
 \end{equation}
 The solutions of Eqs.(\ref{fab0er}) for $f^A_\infty$ and
 $f^B_\infty$ can be used to calculate the stable mutually connected
 giant components of both networks as
\begin{equation}\label{SA}
    \left\{%
\begin{array}{ll}
\mu^A_\infty=p^A_\infty g^A(p^A_\infty)=R^A(1-e^{-c^{BA}_\infty})(1-f^A_\infty)=-\ln({f^A_\infty})/a, \\
\mu^B_\infty= p^B_\infty g^B(p^B_\infty)=R^B(1-e^{-c^{AB}_\infty})(1-f^B_\infty)=-\ln({f^B_\infty})/b.
    \end{array}%
 \right.
\end{equation}
From Eqs.(\ref{SA}), using Eqs.(\ref{cAB}), we can derive the percolation law
for the fractions of the stable giant components of both coupled ER networks

  \begin{eqnarray}
  \mu^A_\infty=R^A(1-e^{-c^{BA}_0\mu^B_\infty})(1-e^{-a\mu^A_\infty}), \label{muSA1} \\
    \mu^B_\infty=R^B(1-e^{-c^{AB}_0\mu^A_\infty})(1-e^{-b\mu^B_\infty}). \label{muSA2} 
 \end{eqnarray}

Eq.(\ref{muSA1}) and Eq.(\ref{muSA2}) are simple and can be related to
the theory of random percolation of a single ER network
\cite{bollobas,bunde}, for which the fraction of the giant component
is $\mu_\infty=R(1-e^{-\langle k\rangle\mu_\infty})$. The coupled ER
networks bring new terms $1-e^{-c^{BA}_0\mu^B_\infty}$ and
$1-e^{-c^{AB}_0\mu^A_\infty}$. In the limit of $c^{AB}_0\to \infty$
(or $c^{BA}_0\to \infty$), the giant component of network B (or
network A) does not depend on the other network and behaves similarly
to the random percolation of a single network.

\begin{figure}
  \includegraphics[width=8cm,angle=-90]{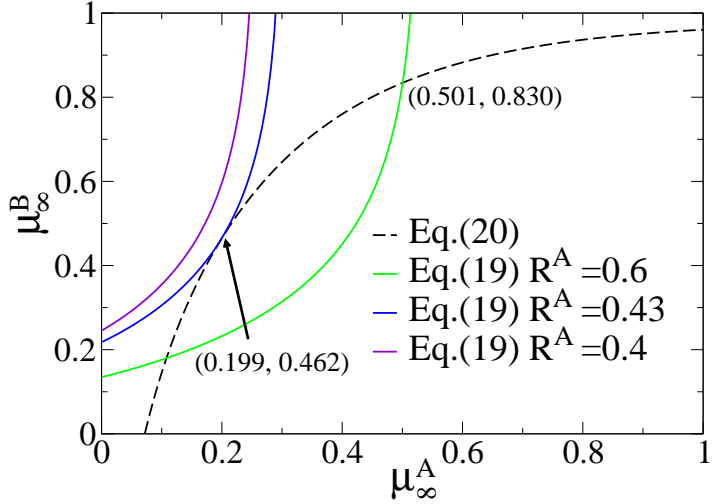}
  \caption{ Demonstration of the functional relation between
    $\mu^B_\infty$ and $\mu^A_\infty$ in Eq.(\ref{muSA1}) and
    Eq.(\ref{muSA2}) for a system of two coupled ER networks with
    $a=b=4$, $c^{AB}_0=c^{BA}_0=4$, and $R^B=1$ with different values
    of $R^A$. Since we use $R^B=1$, at different $R^A$, the relation
    between $\mu^B_\infty$ and $\mu^A_\infty$ given by
    Eq.(\ref{muSA2}) remains the same (shown by the dashed
    line). Eq.(\ref{muSA1}) with $R^A=0.6$, 0.43 and 0.4 are
    shown. One can see that when $R^A<0.43$, there exists no solution
    $(\mu^A_\infty,\mu^B_\infty)$ for Eq.(\ref{muSA1}) and
    Eq.(\ref{muSA2}). The value $1-R^A_{c}\equiv 1-0.43=0.57$
    represents the maximum fraction of nodes in network A one can
    randomly remove at the initial stage of the cascade of failures
    for which the non-zero giant components of both networks still
    exist at the stable state. The abrupt fragmentation of the
    stable giant components at $R^A<R^A_{c}$ represents the first
    order nature of the percolation phase transition.}
\label{fig2}
\end{figure}

From Eqs.(\ref{muSA1}) and (\ref{muSA2}), we find
$\mu^A_\infty$ and $\mu^B_\infty$ for a given set of parameters $R^A$,
$R^B$, $c^{BA}_0$ and $c^{AB}_0$. However, for some values of
$R^A$, $R^B$, $c^{BA}_0$ and $c^{AB}_0$, the solutions for $\mu^A_\infty$
and $\mu^B_\infty$ between 0 and 1 may not exist. There exist critical
  thresholds of $R^A$, $R^B$, $c^{BA}_0$ and $c^{AB}_0$ above
which the two coupled ER networks have non-zero mutually connected
giant components (see Fig.\ref{fig2}).  They are represented as
$R^A_{c}$, $R^B_{c}$, $c^{BA}_{c}$ and $c^{AB}_{c}$.
These values can be solved by finding the tangent point of the two
curves ($\mu^A_\infty$ is plotted as a function of $\mu^B_\infty$ as
shown in Fig.\ref{fig2})
represented by Eqs.(\ref{muSA1}) and (\ref{muSA2}). The thresholds 
  can be found from the tangential condition
\begin{equation}\label{differ}
 \frac{d\mu^A_\infty}{d\mu^B_\infty}\mid_{Eq.(19)}\frac{d\mu^B_\infty}{d\mu^A_\infty}\mid_{Eq.(20)}=1,
\end{equation}
together with Eqs.(\ref{muSA1}) and (\ref{muSA2}).

\section{Numerical Simulations}

\begin{figure}
  \includegraphics[width=6.3cm,angle=-90]{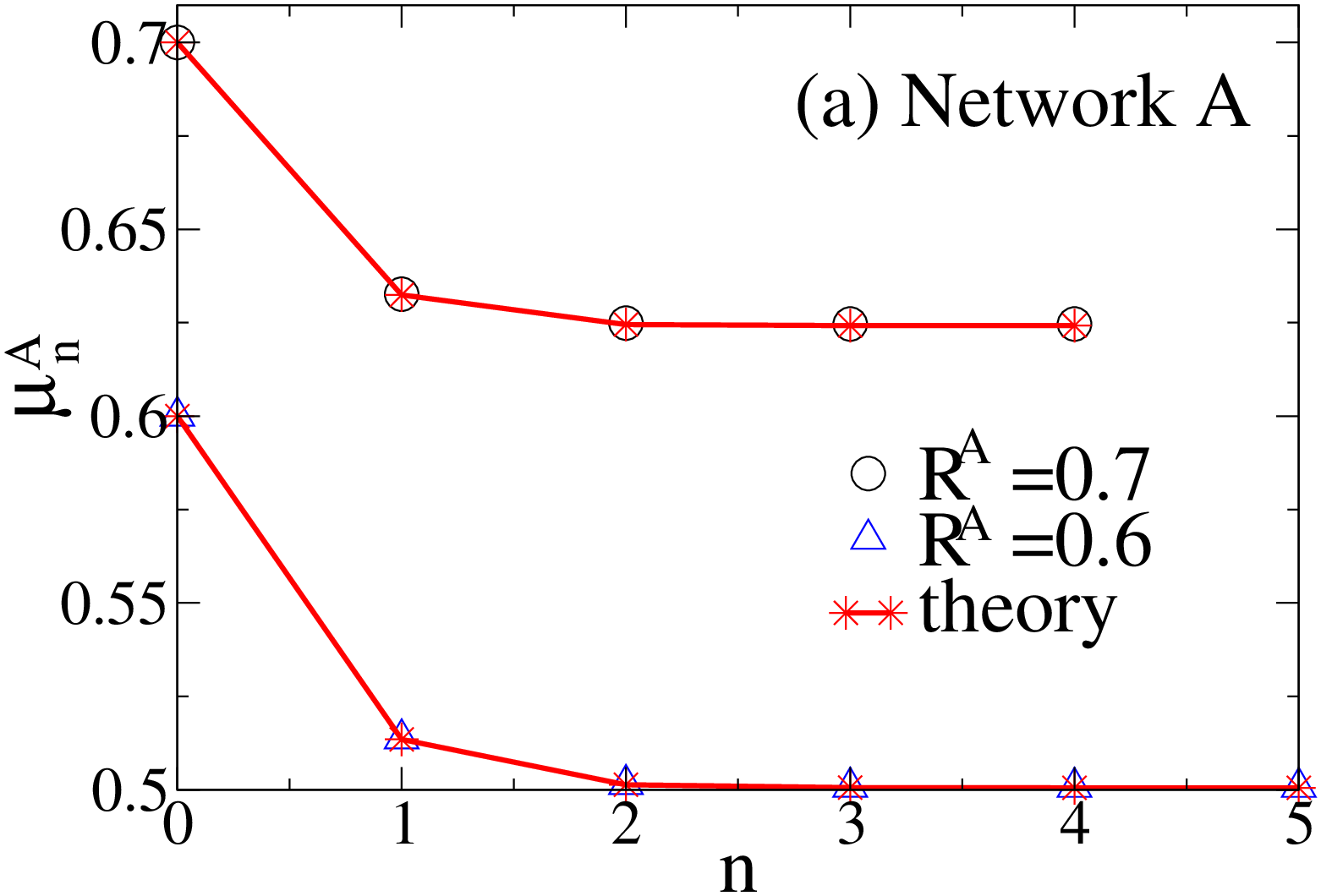}
   \includegraphics[width=6.3cm,angle=-90]{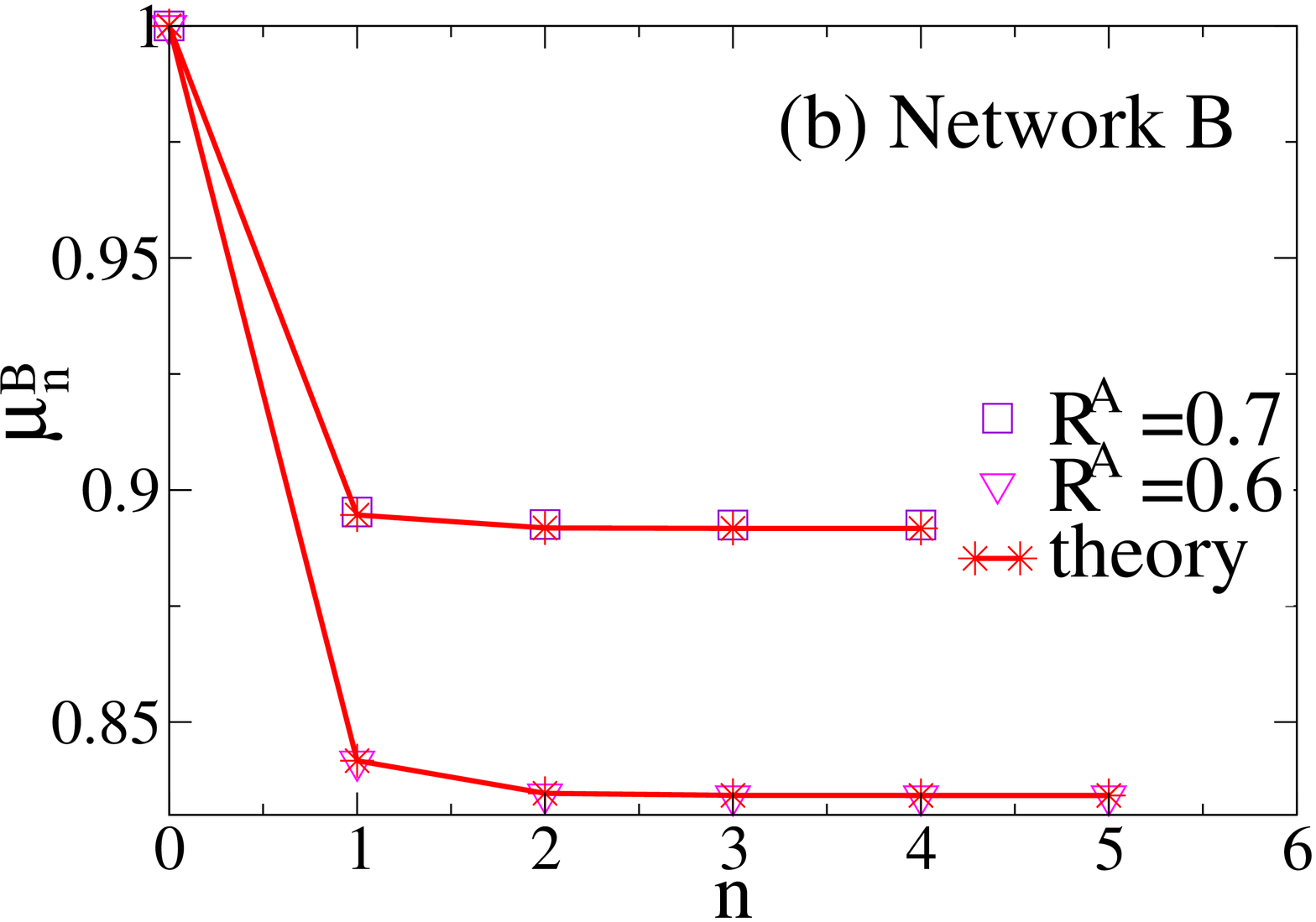}
   \includegraphics[width=6.3cm,angle=-90]{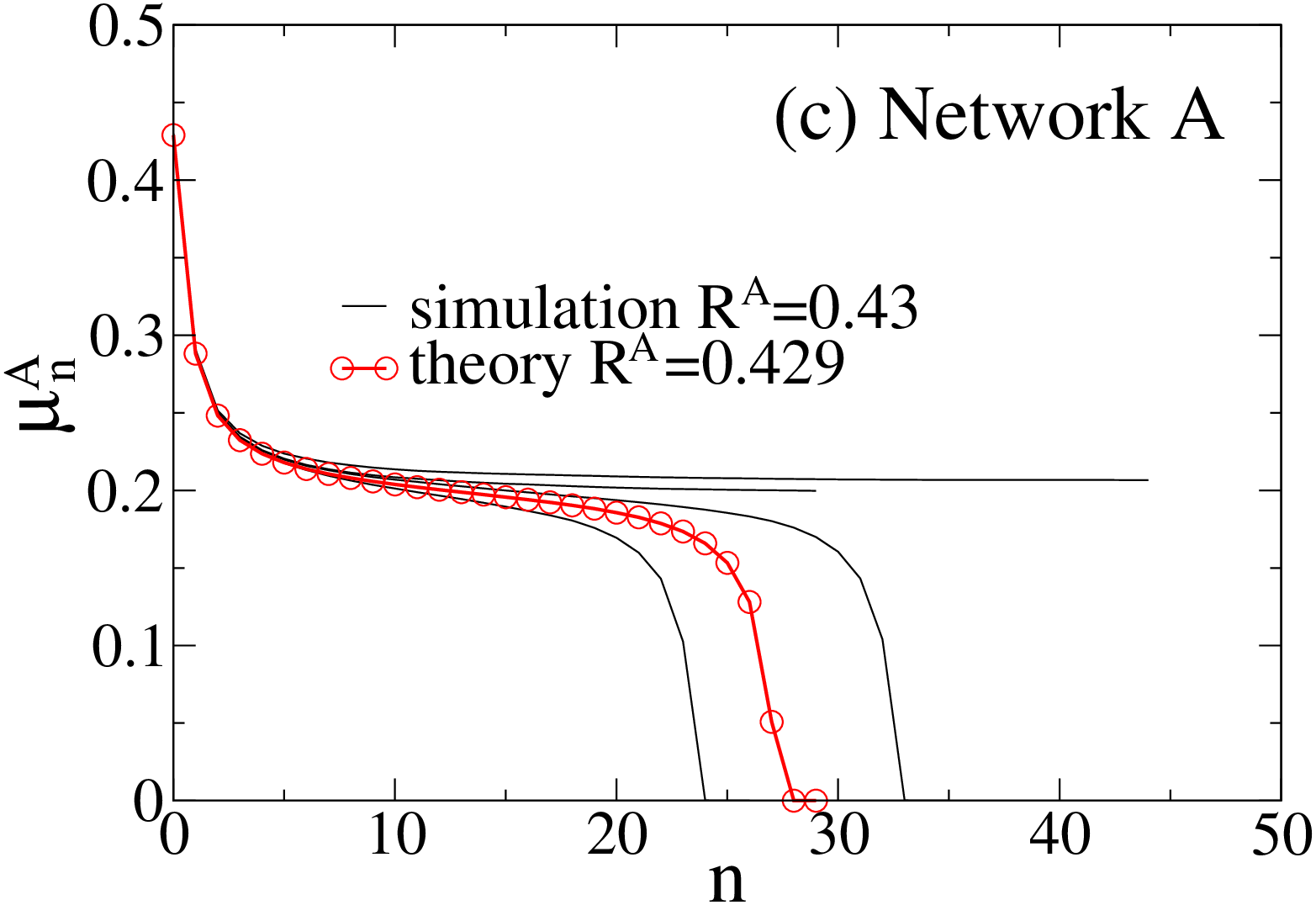}
    \includegraphics[width=6.3cm,angle=-90]{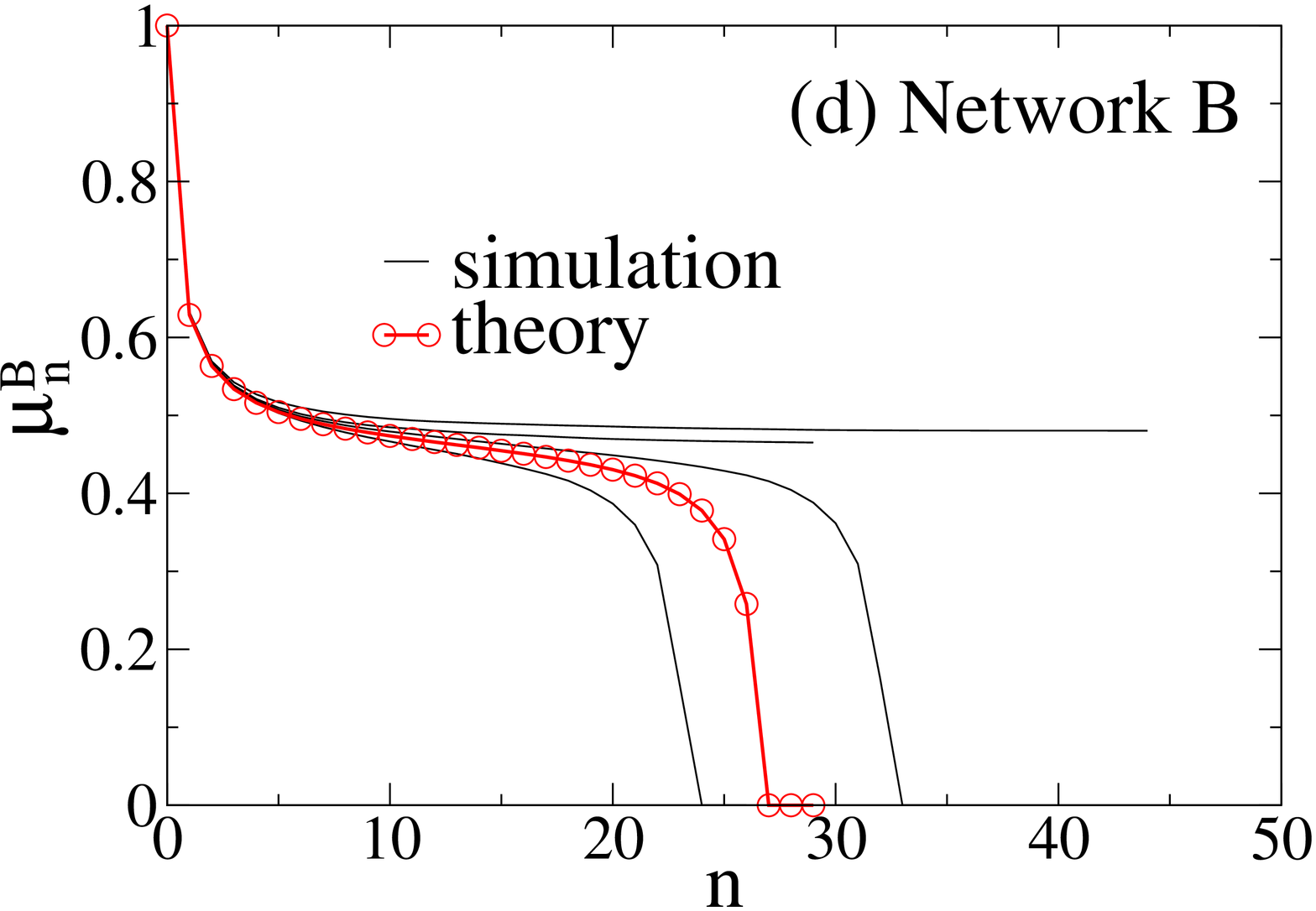}
   \caption{The case of coupled ER networks.
    Comparison between the theoretical predictions, obtained
      from Eqs.(\ref{giABer}), and Eqs.(\ref{muAB}-\ref{cAB}), and
     numerical simulations with $N^A=N^B=10^6$, $a=b=4$, 
     $c^{AB}_0=c^{BA}_0=4$, $R^B=1$ and several values of $R^A$. (a)
     and (b) show $\mu^A_n$ and $\mu^B_n$ at different stages $n$ of
     the cascade of failures for $R^A=0.7$ and 0.6 above $R^A_c\approx
     0.43$ for both theory (lines) and simulations (symbols). One can
     see that both $\mu^A_n$ and $\mu^B_n$ approach a stable value
     $\mu^A_\infty$ and $\mu^B_\infty$ at the end of the cascade of
     failures. The agreement between theory and numerical simulations
     is very good. (c) and (d) show $\mu^A_n$ and $\mu^B_n$ at
       different stages $n$ of the cascade of failures for $R^A\approx R^A_c$.
     The bare lines represent several realizations of the
     simulations and the lines with symbols represent the theoretical
     predictions. One can see that for the early stages (small $n$)
     the agreement is good, however at large $n$ the deviation due to
     random fluctuations in the actual fraction of the giant component
     starts to increase. The random realizations split into two
     classes: one that converges to a non-zero giant component for both
     networks and the other that results in a complete fragmentation.
    }
\label{fig3}
\end{figure}

Next, we compare our theoretical results obtained in Sec. III to
results of numerical simulations. We begin with comparing the
simulations of the stages of the failure cascade in coupled ER
networks with our theoretical predictions. In all our simulations, we
use $N^A=N^B=10^6$.  Fig. \ref{fig3} shows $\mu^A_n$ and $\mu^B_n$ as a
function of $n$ for $a=b=4$,
$c^{AB}_0=c^{BA}_0=4$, $R^B=1$ and for different values of $R^A$. One
sees very good agreement between the theory and the
simulations. Close to $R^A_{c}$, both $\mu^A_n$ and $\mu^B_n$ show
large fluctuations between different realizations (shown in Figs. 3c
and 3d). The random realizations split into two classes: one that
converges to a non-zero giant component for both networks and the
other that results in a complete fragmentation. The agreement between
the simulations and theoretical predictions is also good for
different values of $R^B$, $a$, $b$ and $c^{AB}_0$ and $c^{BA}_0$.



\begin{figure}
  \includegraphics[width=6.3cm,angle=-90]{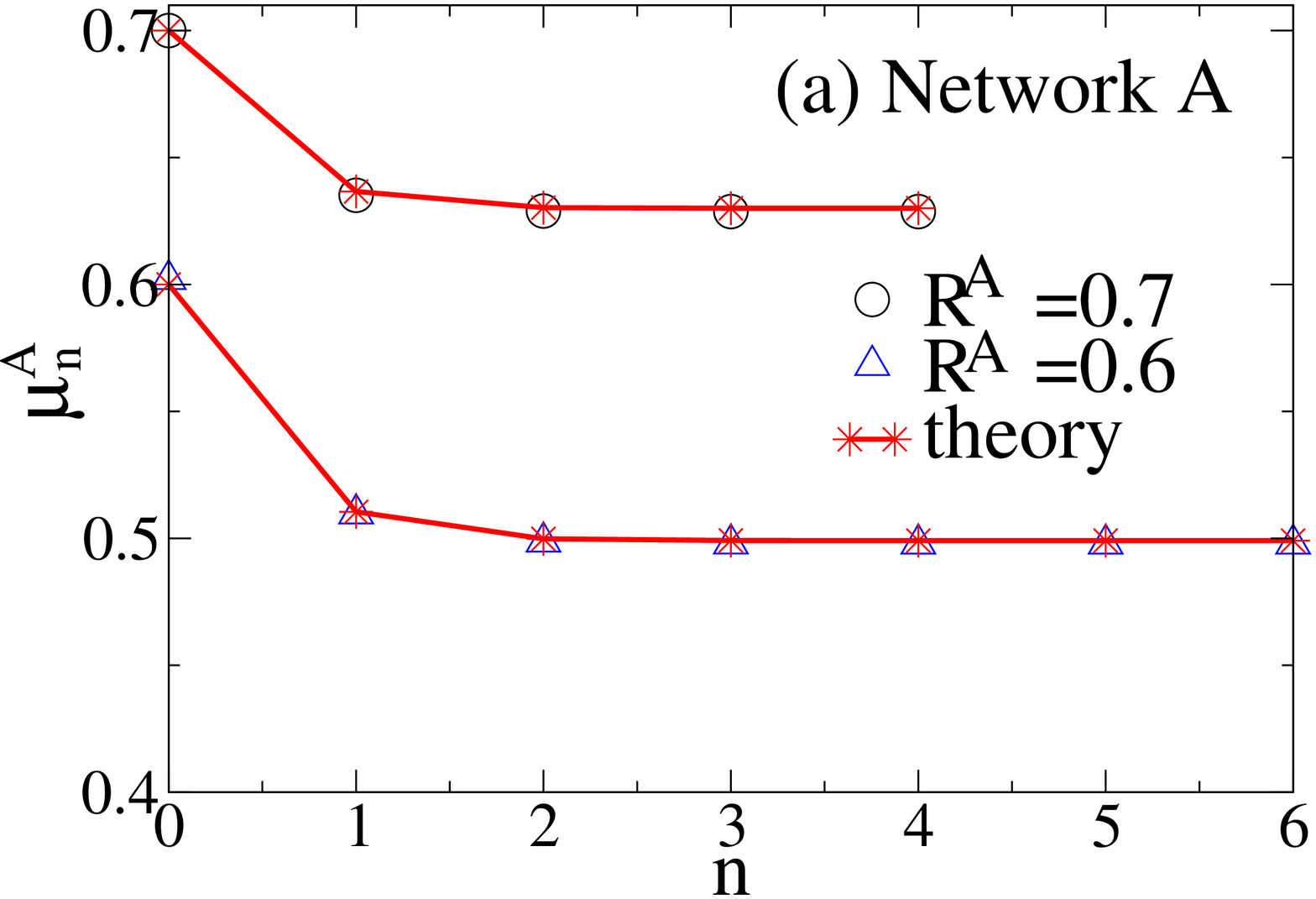}
   \includegraphics[width=6.3cm,angle=-90]{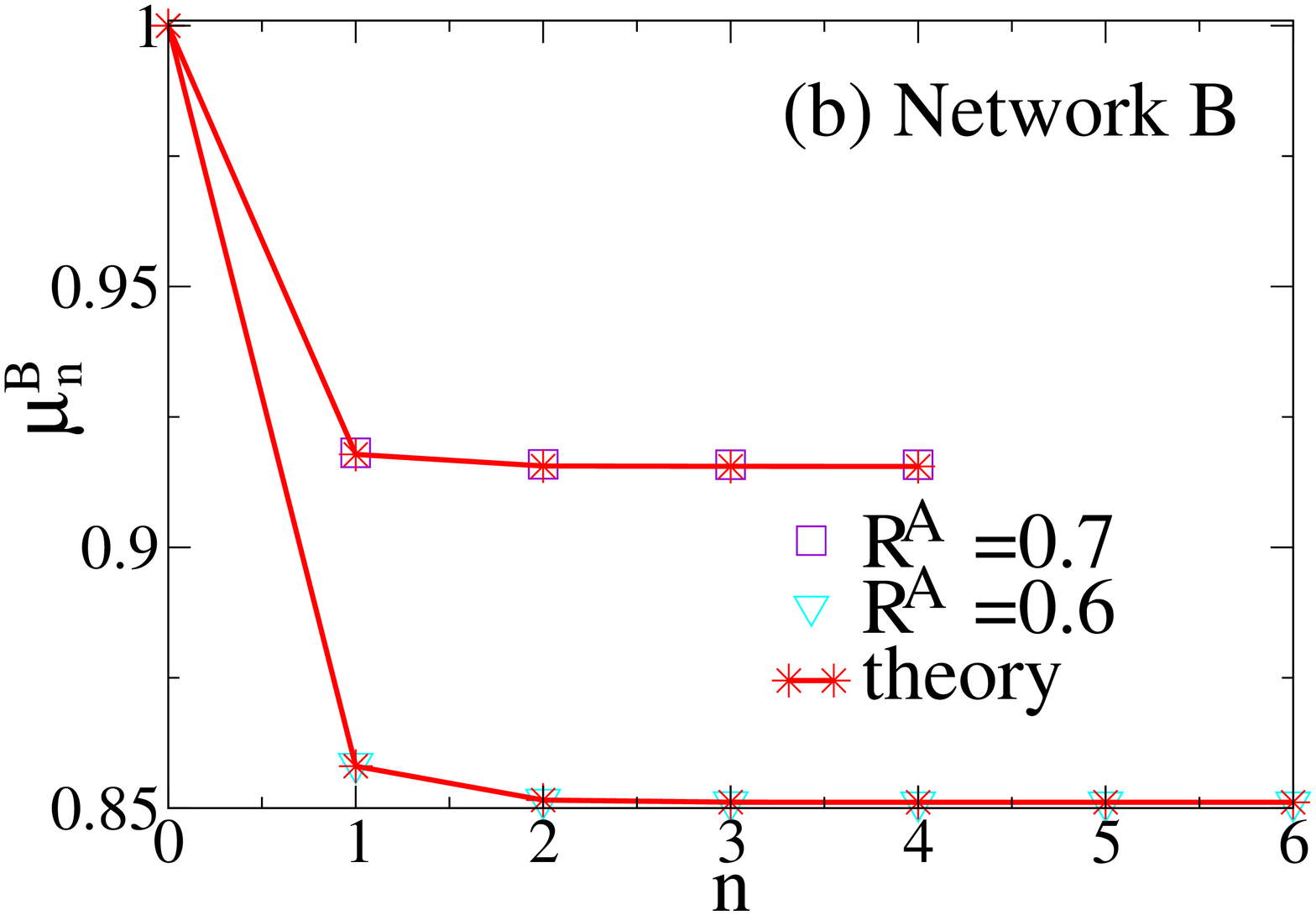}
   \includegraphics[width=6.3cm,angle=-90]{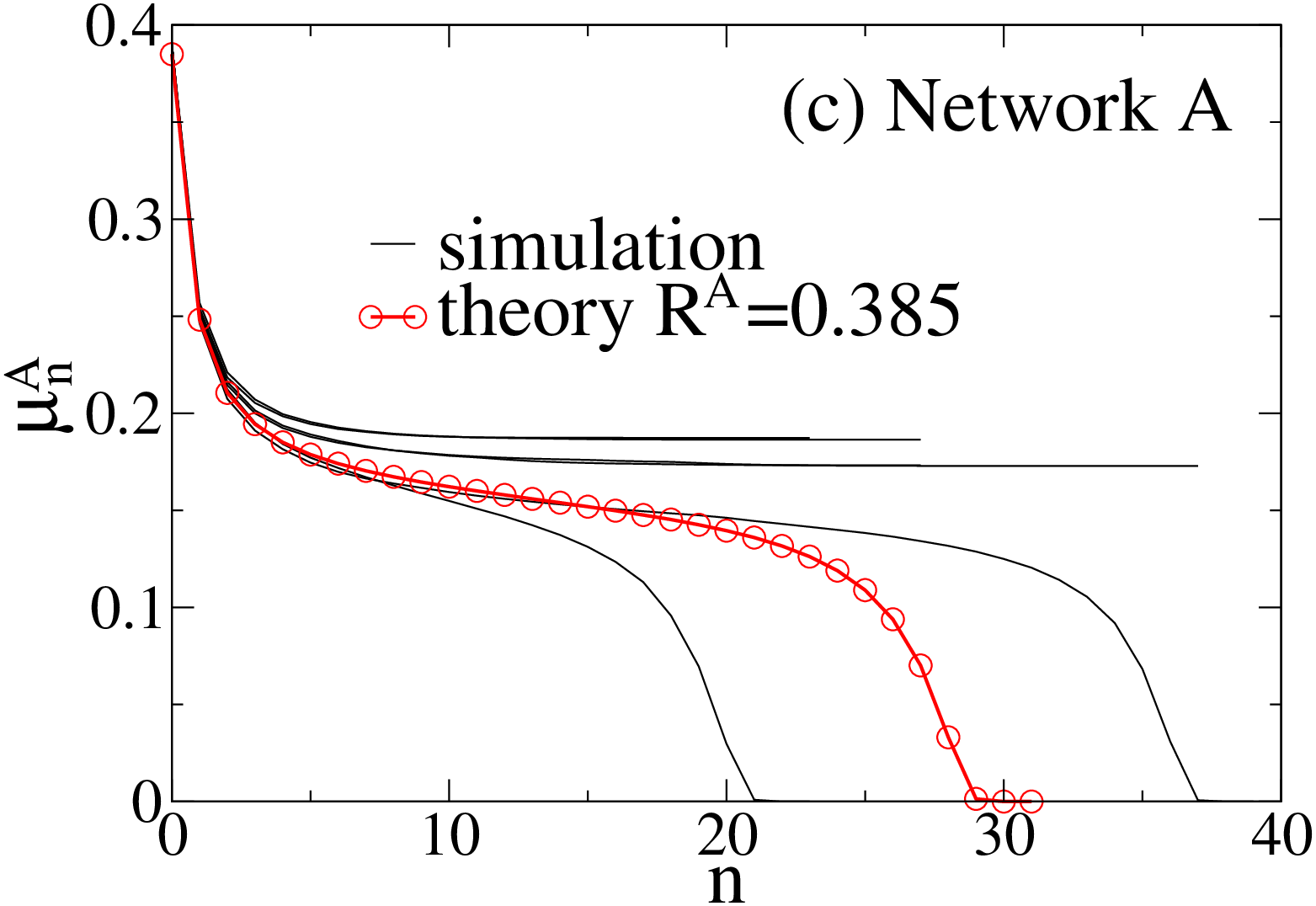}
    \includegraphics[width=6.3cm,angle=-90]{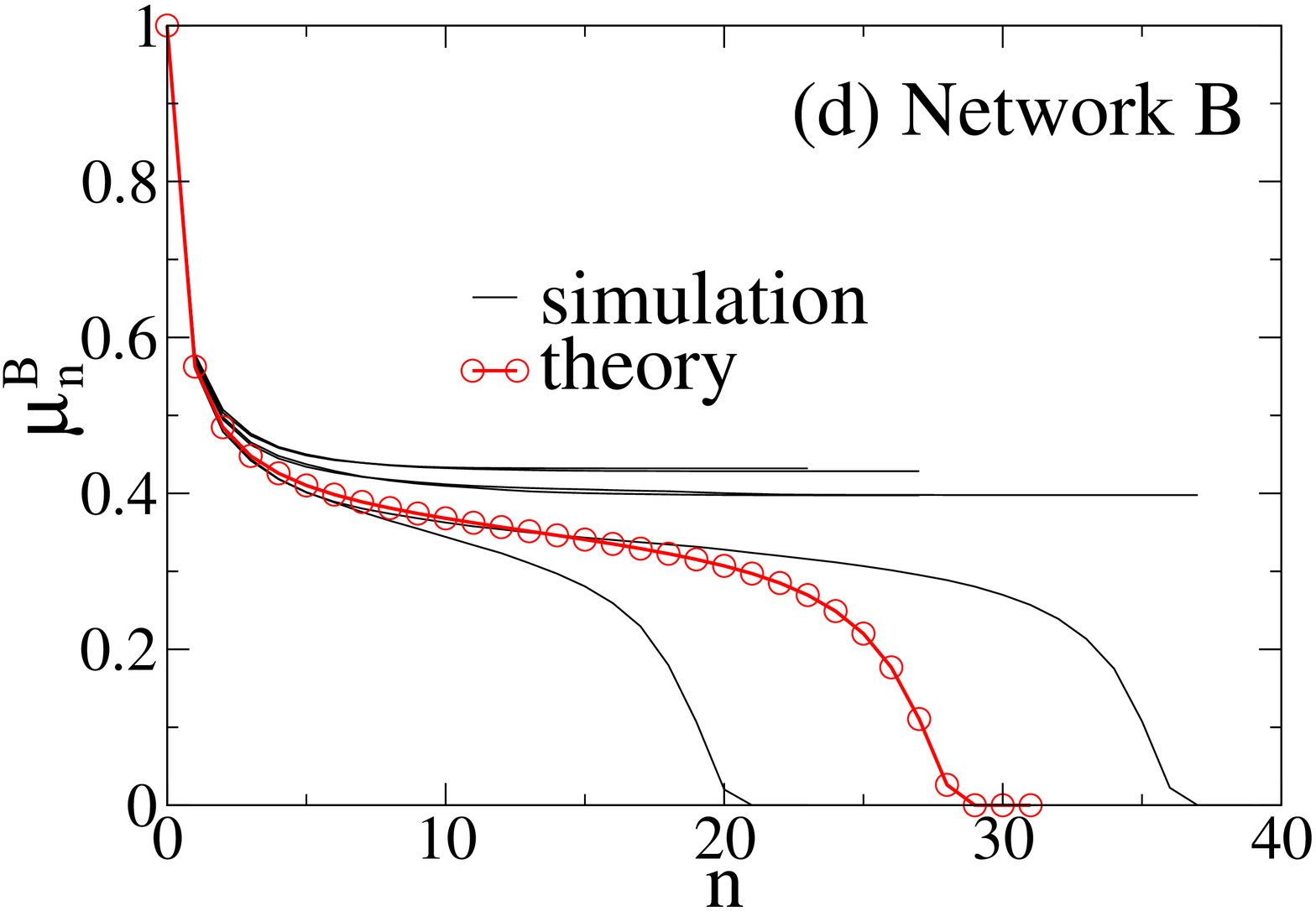}
   \caption{The case of coupled SF networks.
     Comparison between the theoretical predictions, obtained
     from Eqs.(\ref{giAB}), and Eqs.(\ref{fAB}-\ref{cAB}), and
     numerical simulations with $N^A=N^B=10^6$,
     $\lambda^A=\lambda^B=2.5$, $c^{AB}_0=c^{BA}_0=4$, $R^B=1$ and
     different values of $R^A$. (a) and (b) show $\mu^A_n$ and
     $\mu^B_n$ at different stages $n$ of the cascade of failures for
     $R^A=0.7$ and 0.6 above $R^A_c\approx 0.385$ for both theory
     (lines) and simulations (symbols). Similar to Fig. \ref{fig3}, one
     can see that both $\mu^A_n$ and $\mu^B_n$ approach a stable value
     $\mu^A_\infty$ and $\mu^B_\infty$ at the end of cascade
     failures. The agreement between the theory and numerical
     simulations is very good. (c) and (d) show $\mu^A_n$ and
       $\mu^B_n$ at different stages $n$ of the cascade of failures
       for $R^A\approx R^A_c$.
     Bare lines represent several realizations of the simulations and
     the lines with symbols represent the theoretical predictions. One
     can see that for the early stages the agreement is good, however
     at large $n$ the deviation due to random fluctuations in the
     actual fraction of the giant component increase. The random
     realizations split into two classes: one that converges to a
     non-zero giant component for both networks and the other that
     results in a complete fragmentation.  For coupled SF networks, at
     $R^A=R^A_{c}$, the fluctuations of both $\mu^A_n$ and $\mu^B_n$
     seem to be relatively larger than that of coupled ER networks,
     due to the existence of large degree nodes in SF networks.
    }
\label{fig4}
\end{figure}

In Fig. \ref{fig4}, we compare the theoretical predictions and
simulations of the giant components at
stages of the cascade of failures for a system of two
coupled SF networks with $\lambda^A=\lambda^B=2.5$,
$c^{AB}_0=c^{BA}_0=4$, $R^B=1$ and different values of
$R^A$. Similarly, we obtain agreement between the theoretical predictions
and the simulations. Close to $R^A_{c}$, both
$\mu^A_n$ and $\mu^B_n$ of different realizations show large
fluctuations and the random realizations also split into two classes.
We also simulated other values of $R^B$, $\lambda^A$, $\lambda^B$,
$c^{AB}_0$ and $c^{BA}_0$ and found very good agreement between and
theoretical predictions and simulations.

\begin{figure}
  \includegraphics[width=6.3cm,angle=-90]{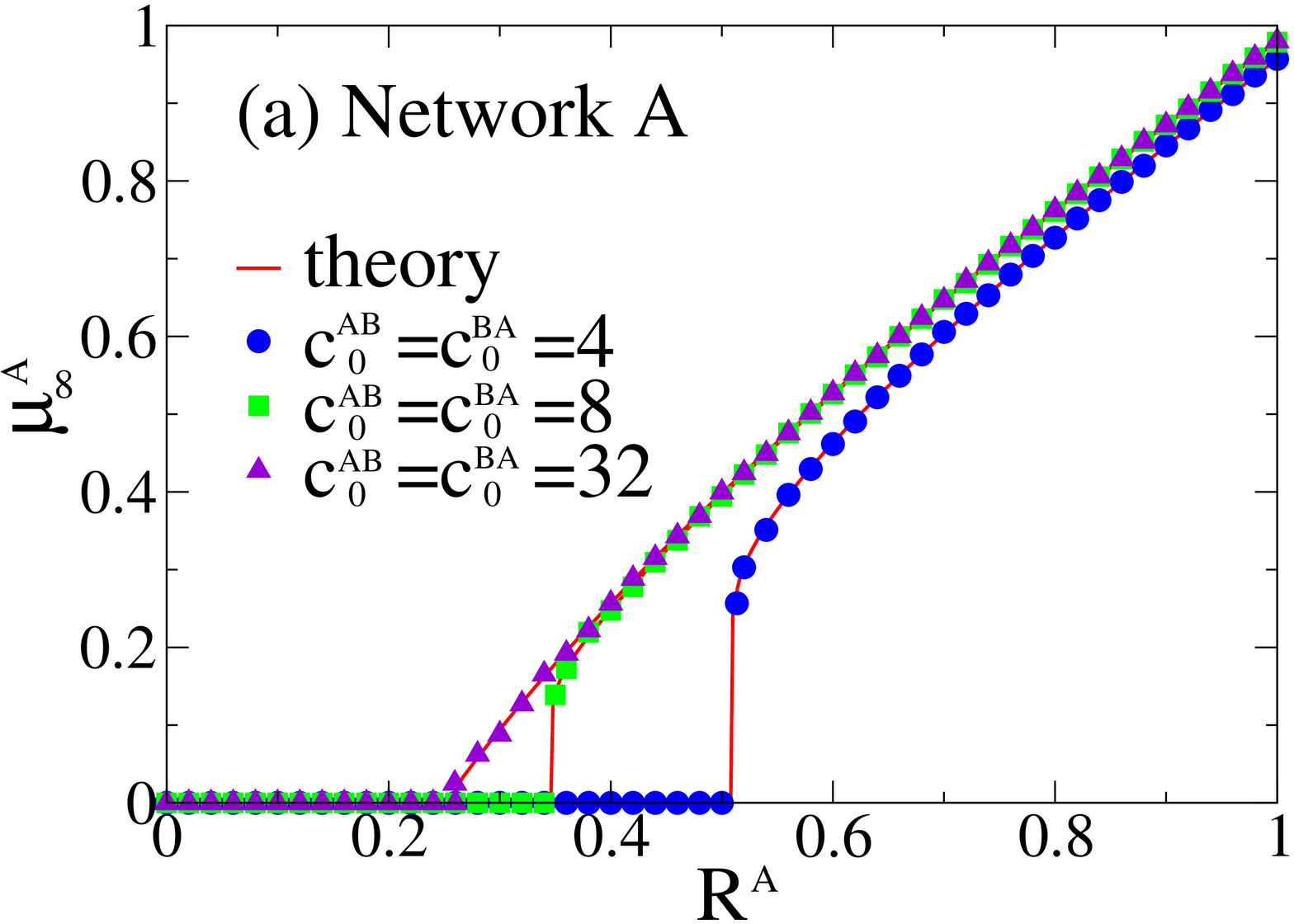}
   \includegraphics[width=6.3cm,angle=-90]{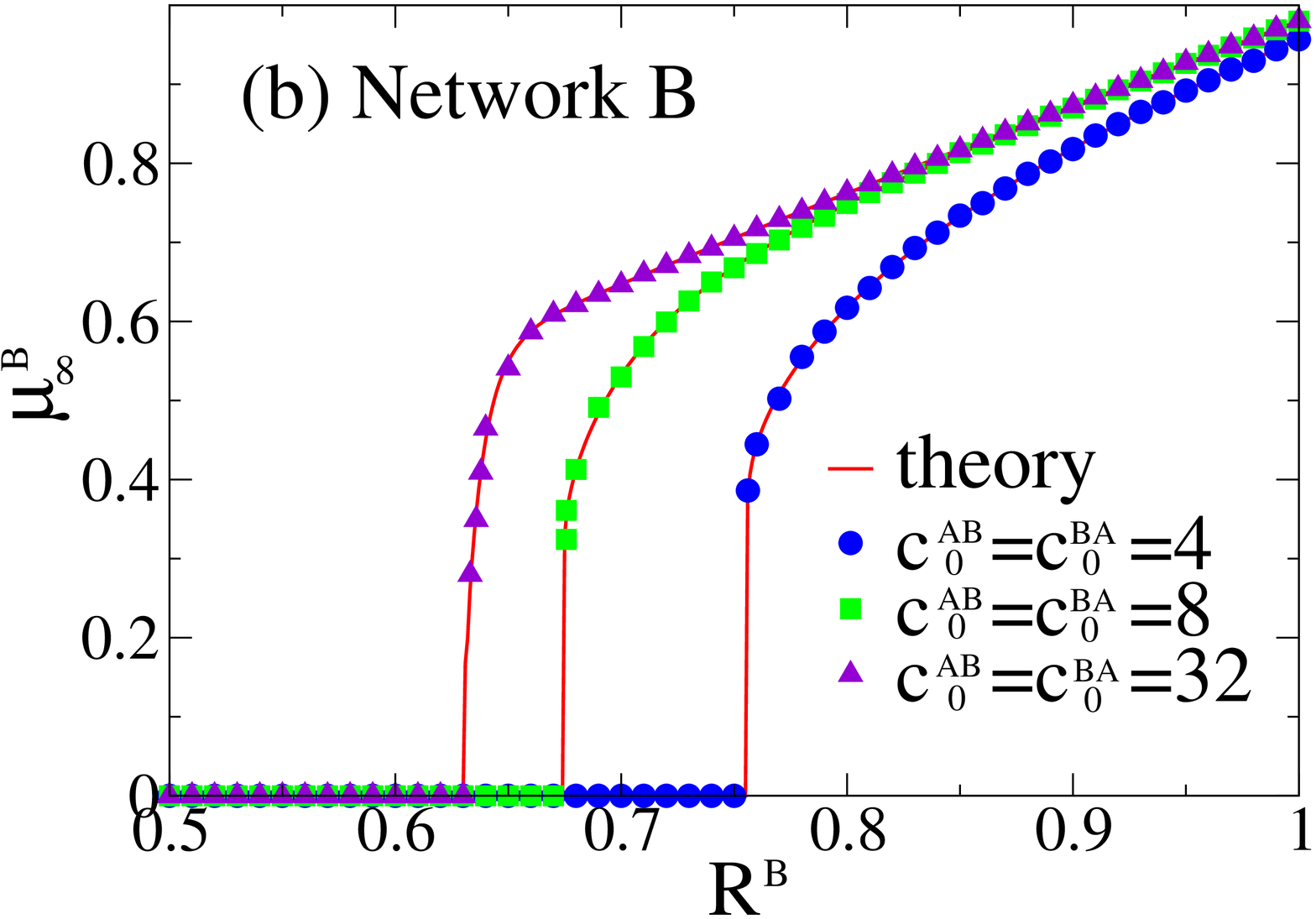}
 \caption{ (a) $\mu^A_\infty$ and (b) $\mu^B_\infty$ in the stable
   states as a function of $R^A$ and $R^B$ for coupled ER networks A
   and B with $N^A=N^B=10^6$, $a=b=4$, $c^{AB}_0=c^{BA}_0=4$ and
   $1-R^A=2(1-R^B)$. Several curves for $c^{AB}_0=c^{BA}_0=4$, 8 and
   32 are shown. The theory (lines) fits very well with the simulation
   results (symbols). One can see that for a given set of $c^{AB}_0$
   and $c^{BA}_0$ there exist critical thresholds $R^A_{c}$ and
   $R^B_{c}$, below which both networks will collapse and have no
   stable non-zero giant components. The value of $R^A_{c}$ approaches
   the critical threshold of random percolation ($r=1/a=0.25$) of a
   single network for large values of $c^{AB}_0$ and $c^{BA}_0$. The
   initial attack on network B is smaller than that on network A and
   thus $R^B_{c}>R^A_{c}$.  }
\label{fig5}
\end{figure}

The fractions of the giant components of both network A
($\mu^A_\infty$) and network B ($\mu^A_\infty$) in the stable state
can be found from Eqs.(\ref{muSA1}) and (\ref{muSA2}) for coupled ER
networks. We solve these equations numerically for different $R^A$ and
$R^B$, and compare the theoretical predictions with the simulation
results (Fig. \ref{fig5}). For simplicity, we assume $a=b=4$
and that the initial fraction of nodes affected by the random attack
in network A is twice as large as that in network B
($1-R^A=2(1-R^B)$). We test different average degrees of inter-links for both
networks $c^{AB}_0=c^{BA}_0$.

In Fig. \ref{fig5}, we present results for the giant components of
both networks as a function of $R^A$ and $R^B$.  For different sets of
$c^{AB}_0$ and $c^{BA}_0$, we find again that the theory fits well
with simulation results. One can see the critical $R^A_{c}$ and
$R^B_{c}$, which are the minimum fractions of both networks needed to
be kept at the beginning of the cascade of failures in order to have
non-zero connected giant components of both networks at the stable
state. At $R^A_{c}$ and $R^B_{c}$, both $\mu^A_\infty$ and
$\mu^B_\infty$ show an abrupt change from a finite fraction ($\mu^A_c$
and $\mu^B_c$) to zero. As $c^{AB}_0$ and $c^{BA}_0$ increase,
$R^A_{c}$ approaches the critical threshold of random percolation of a
single ER network, which is $1/a$. As expected for single networks,
$\mu^A_c$ and $\mu^B_c$ approach 0 and a second order phase transition
exists for large $c^{AB}_0$ and $c^{BA}_0$. However, for finite
$c^{AB}_0$ and $c^{BA}_0$ the changes of $\mu^A_\infty$ and
$\mu^B_\infty$ are not continuous at $R^A_{c}$ and $R^B_{c}$,
indicating a first order phase transition.  This result is predicted
by Eqs.(\ref{muSA1}) and (\ref{muSA2}). We find that the theory fits
well with the simulation results for the entire range of $R^A_{c}$ and
$R^B_{c}$ for different values of $c^{AB}_0$ and $c^{BA}_0$.

\begin{figure}
  \includegraphics[width=6.3cm,angle=-90]{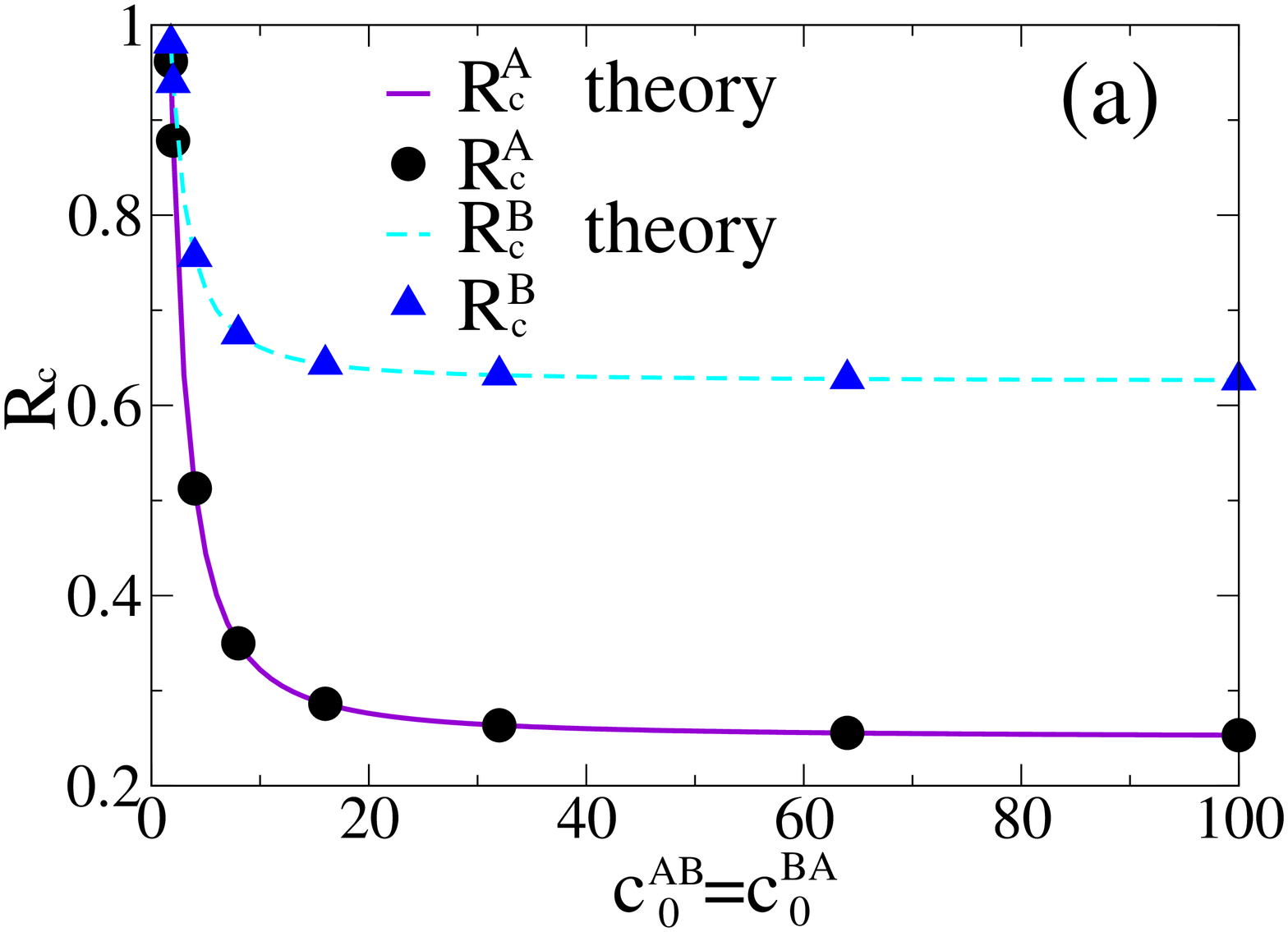}
   \includegraphics[width=6.3cm,angle=-90]{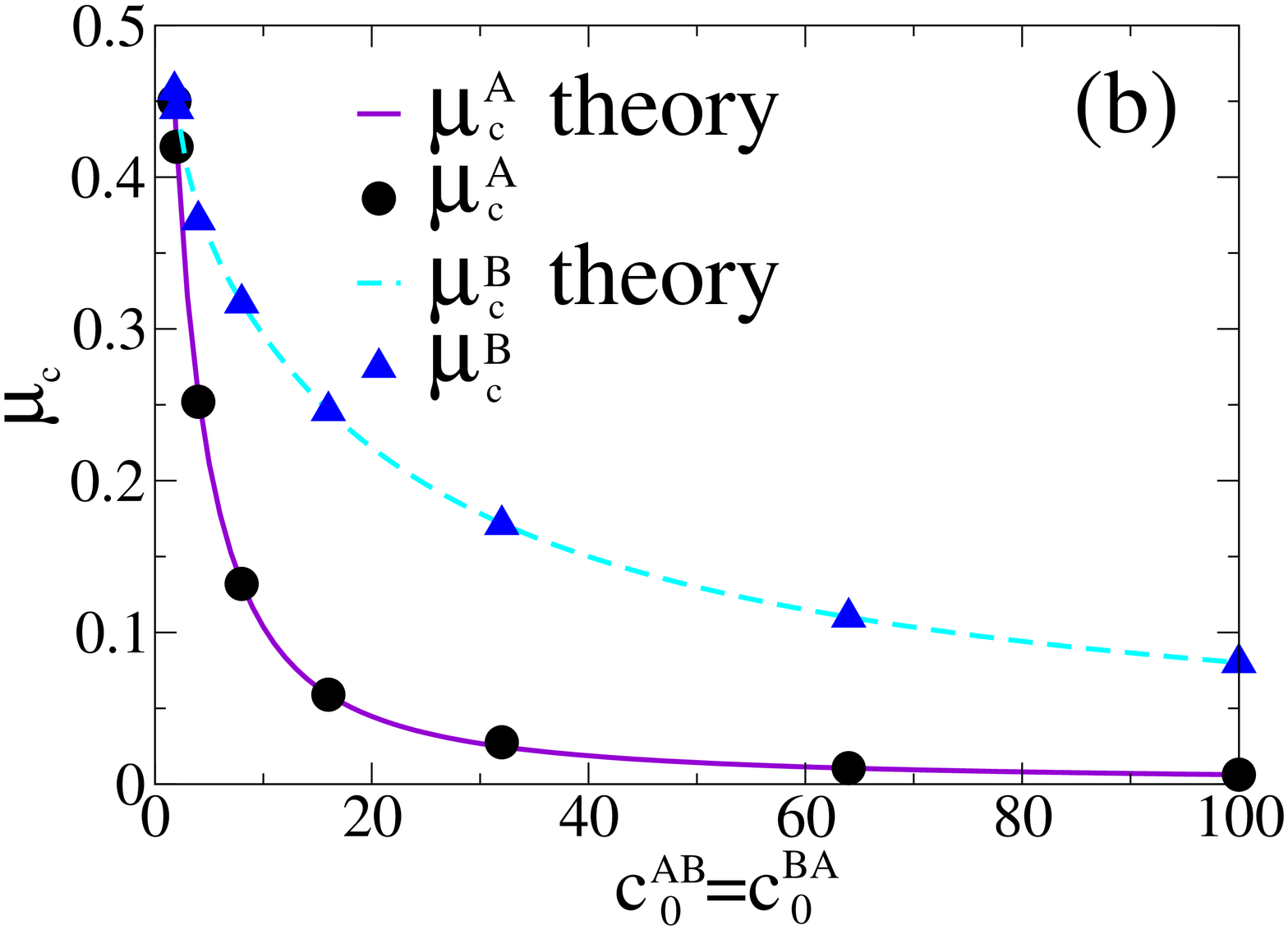}
  \caption{The dependences of $R_{c}$ (a) and $\mu_{c}$ (B) on
    $c^{AB}_0=c^{BA}_0$ for coupled ER networks with $N^A=N^B=10^6$,
    $a=b=4$, and $1-R^A=2(1-R^B)$. In (a), the critical initial
    fraction of the network A, $R^A_{c}$, and the critical initial
    fraction of the network B, $R^B_{c}$, are shown as a function of
    $c^{AB}_0=c^{BA}_0$. The theory (full line and dashed lines) fits
    well the simulation results (symbols). $R^A_{c}$ approaches the
    critical threshold ($1/a$) of random percolation for a single
    network, as predicted by Eqs.(\ref{muSA1}) and (\ref{muSA2}). In
    (b), the giant component of both network $\mu^A_{c}$ and
    $\mu^B_{c}$ are shown as a function of $c^{AB}_0=c^{BA}_0$ at
    $R^A_{c}$ and $R^B_{c}$.  The theory (full line and dashed lines)
    fits well the simulation results (symbols). One can see that for
    large $c^{AB}_0=c^{BA}_0$, $\mu^A_{c}$ and $\mu^B_{c}$ both
    approach zero as expected for a single network. However,
    $\mu^A_{c}$ and $\mu^B_{c}$ will never reach zero for finite
    $c^{AB}_0$ and $c^{BA}_0$, the phase transition thus remain a
    first order. }
\label{fig6}
\end{figure}

Next, we study the dependence of $R^A_{c}$ and $R^B_{c}$, $\mu^A_c$
and $\mu^B_c$ on $c^{AB}_0$ and $c^{BA}_0$ (see Fig.\ref{fig6}). For
simplicity and for comparing with our earlier cases, we use the same
set of parameter for both network: $a=b=4$, $c^{AB}_0=c^{BA}_0$, and
$1-R^A=2(1-R^B)$. As seen from Fig.\ref{fig6}, the theory fit well the
numerical simulations. For large $c^{AB}_0=c^{BA}_0$, one can see that
$R^A_{c}$ approaches the random percolation threshold $1/a$ on a
single ER network. This behavior indicates that when one network A has
enough support from network B and {\it vice versa}, both networks will
behave as if they are independent. Indeed for large $c^{AB}_0$ and
$c^{BA}_0$, at $R^A_{c}$ and $R^B_{c}$, the stable giant components of
both networks $\mu^A_c$ and $\mu^A_c$ approach zero as expected for a
second order percolation phase transition. However, as seen from
Eqs.(\ref{muSA1}) and (\ref{muSA2}), for finite values of $c^{AB}_0$
and $c^{BA}_0$, neither $\mu^A_c$ nor $\mu^B_c$ is zero. This result
supports the existence of a first order phase transition for the
entire range of $c^{AB}_0$ and $c^{BA}_0$.  Good agreement between
theory and simulations, and similar behavior of $R^A_{c}$, $R^B_{c}$,
$\mu^A_c$ and $\mu^A_c$ as a functions of $c^{AB}_0$ and $c^{BA}_0$
have been found for other sets of parameters.

\section{Conclusions and discussions}
In this paper, we extend previous works \cite{buldyrev,parshani} on
the cascade of failures on interdependent networks by considering
random support-dependent relations between two coupled network
systems. Our theory is in excellent agreement with the numerical
simulations on coupled Erd\H{o}s-R\'enyi (ER) and coupled scale-free
(SF) networks systems. For coupled ER networks, the percolation law
for the giant components of both networks have a simple form, which in
the limit of large number of supports gives the percolation law for
single networks.  Only in the limit of a large number of support is
the percolation transition of second order, while in general case, the
coupled network show a first order phase transition.  Our model can
help to further understand real-life coupled network systems, where
complex dependence-support relations exists.  Recently, a
complementary approach to study the robustness of coupled networks
system has been proposed \cite{leicht}, which is based on a quite
different assumption about the way networks are coupled.  In contrast
to our case where $p_c$ increases due to coupling, in their case $p_c$
decreases.  Note that there are also recent efforts to study the
robustness of single networks \cite{gallos,moreiral,hooyberghs}
undergoing targeted percolation, which correlates with the topology of
the network. In the same spirit, our work can be extended to study the
robustness of coupled networks under non-random percolation. A first
attempt in this direction for interdependent networks can be found in
Ref. \cite{huang}.

 
\begin{acknowledgements}

We wish to thank the ONR, DTRA, EU project Epiwork, and the Israel
Science Foundation for financial support.  S.V.B. thanks the Office of
the Academic Affairs of Yeshiva University for funding the Yeshiva
University high performance computer cluster and acknowledges the
partial support of this research through the Dr.  Bernard W. Gamson
Computational Science Center at Yeshiva College.
\end{acknowledgements}

\vspace*{-0.3cm}


\end{document}